\documentclass[twocolumn,showpacs,preprintnumbers,amsmath,amssymb,prb,aps]{revtex4}

\usepackage{epsfig}
\usepackage{graphicx}
\usepackage{dcolumn}
\usepackage{bm}

\begin{document}
\title{Importance of Co 3$d$ electron correlation in a Ce-based Kondo
lattice, Ce$_2$CoSi$_3$}
\author{Swapnil Patil}
\author{Sudhir Pandey}
\author{V. R. R. Medicherla}
\author{R. S. Singh}
\author{R. Bindu}
\author{E. V. Sampathkumaran}
\author{Kalobaran Maiti}
\altaffiliation{Electronic mail: kbmaiti@tifr.res.in}

\affiliation{Department of Condensed Matter Physics and Materials
Science, Tata Institute of Fundamental Research, Homi Bhabha Road,
Colaba, Mumbai 400005, India}
\date{\today}
\begin{abstract}

We study the role of electron correlations among Co 3$d$ electrons
contributing to the conduction band of a Kondo lattice compound,
Ce$_2$CoSi$_3$, using high resolution photoemission spectroscopy and
{\it ab initio} band structure calculations. Experimental results
reveal signature of Ce 4$f$ states derived Kondo resonance feature
at the Fermi level and dominance of Co 3$d$ contributions at higher
binding energies in the valence band. The line shape of the
experimental Co 3$d$ band is found to be significantly different
from that obtained from the band structure calculations within the
local density approximations. Consideration of electron-electron
Coulomb repulsion among Co 3$d$ electrons leads to a better
representation of experimental results. The correlation strength
among Co 3$d$ electrons is found to be about 3 eV. Signature of
electron correlation induced satellite feature is also observed in
the Co 2$p$ core level spectrum. Thus, these results demonstrate the
importance of the electron correlation among conduction electrons to
derive the microscopic description of such Kondo systems.

\end{abstract}
\maketitle

\section{Introduction}

Study of Ce-intermetallics have drawn significant attention during
past few decades due to the observation of many unusual properties
such as valence fluctuations, Kondo screening, heavy fermion
superconductivity in these systems. Such properties arise due to the
proximity of Ce 4$f$ level to the Fermi level leading to strong
hybridization between the Ce 4$f$ states and the conduction
electronic states.\cite{Brandt} A lot of success has been achieved
to describe these systems within the Anderson impurity models. Here,
the parameters defining the hybridization between 4$f$ states and
valence electronic states are often estimated using band structure
calculations based on local density approximations
(LDA).\cite{Willis,Gunnarsson1,Gunnarsson2,Gunnarsson3} However, the
scenario can be significantly different if these materials contain
transition metals; the $d$ electrons contribute to the conduction
band and the finite correlation strength among them leads to
non-applicability of the band structure results within LDA.

Here, we report the results of our investigations on the electronic
structure of Ce$_2$CoSi$_3$ using high resolution photoemission
spectroscopy and {\it ab initio} band structure calculations.
Ce$_2$CoSi$_3$ crystallizes in a AlB$_2$ derived hexagonal structure
(space group $P6/mmm$) and is a mixed valent (Kondo lattice)
compound.\cite{Gordon,Majumdar,Patil1} The electrical transport
measurements revealed temperature dependence\cite{Patil1} typical of
a mixed valent system.\cite{Lawrence} No signature of magnetic
ordering was observed in the magnetic susceptibility measurements
down to 0.5 K.\cite{Patil1} Interestingly, gradual substitution of
Rh at Co sites leads to plethora of interesting features due to
increasing dominance of indirect exchange interaction.\cite{Patil1}
For example, $x$ = 0.6 composition in Ce$_2$Rh$_{1-x}$Co$_x$Si$_3$
exhibits quantum critical behavior. Intermediate compositions having
higher Rh concentration exhibit signature of spin density wave (SDW)
state.\cite{Patil1} Thus, the $d$ electronic states corresponding to
the transition metals plays a key role in determining the electronic
properties in this interesting class of compounds. Therefore,
Ce$_2$CoSi$_3$ could be one of the ideal ones where the correlation
among the electronic states other than Ce 4$f$ states is important.

High energy resolution employed in our measurements enabled us to
reveal Kondo-resonance feature and the corresponding spin orbit
satellite. The comparison of the experimental spectra and the
calculated ones indicate that the correlation strength among Co 3$d$
electrons is strong ($\sim$ 3 eV). The contribution of the Co 3$d$
partial density of states (PDOS) is small in the vicinity of the
Fermi level, where Ce 4$f$ contributions are dominant.

\section{Experimental details}

Ce$_2$CoSi$_3$ was prepared by melting together stoichiometric
amounts of high purity ($>$ 99.9\%) Ce, Co and Si in an arc furnace.
The single phase was confirmed by the absence of impurity peaks in
the $x$-ray diffraction pattern. The specimen was further
characterized by scanning electron microscopic measurements and
energy dispersive $x$-ray analysis.\cite{Patil1} The photoemission
measurements were performed using a Gammadata Scienta SES2002
analyzer and monochromatic laboratory photon sources. The energy
resolutions were set to 0.4 eV, 5 meV and 5 meV at Al $K\alpha$
(1486.6 eV), He {\scriptsize II}$\alpha$ (40.8 eV) and He
{\scriptsize I}$\alpha$ (21.2 eV) photon energies, respectively. The
base pressure in the vacuum chamber was 3 $\times$ 10$^{-11}$ torr.
The temperatures variation down to 20 K was achieved by an open
cycle He cryostat (LT-3M, Advanced Research Systems, USA). The
sample surface was cleaned by {\it in situ} scraping using a diamond
file and the surface cleanliness was ensured by the absence of O
1$s$ and C 1$s$ features in the $x$-ray photoelectron (XP) spectra
and the absence of impurity features in the binding energy range of
5-6 eV in the ultraviolet photoelectron (UP) spectra. The
reproducibility of the spectra was confirmed after each trial of
cleaning process.

\section{Calculational details}

The electronic band structure of Ce$_2$CoSi$_3$ was calculated using
{\it state-of-the-art} full potential linearized augmented plane
wave (FLAPW) method using WIEN2k software\cite{wien} within the
local density approximations, LDA. The convergence for different
calculations were achieved considering 512 $k$ points within the
first Brillouin zone. The error bar for the energy convergence was
set to $<$~0.2~meV per formula unit (fu). In every case, the charge
convergence was achieved to be less than 10$^{-3}$ electronic
charge. The lattice constants used in these calculations are
determined from the $x$-ray diffraction patterns considering AlB$_2$
derived hexagonal structure and are found to be $a$~=~8.104~\AA\ and
$c$ =~4.197~\AA.\cite{Gordon} The muffin-tin radii ($R_{MT}$) for
Ce, Co and Si were set to 2.5 a.u., 2.28 a.u. and 2.02 a.u.,
respectively.

\section{Results and discussions}

\begin{figure}
 \vspace{-2ex}
 \begin{center}
\includegraphics [scale=0.45]{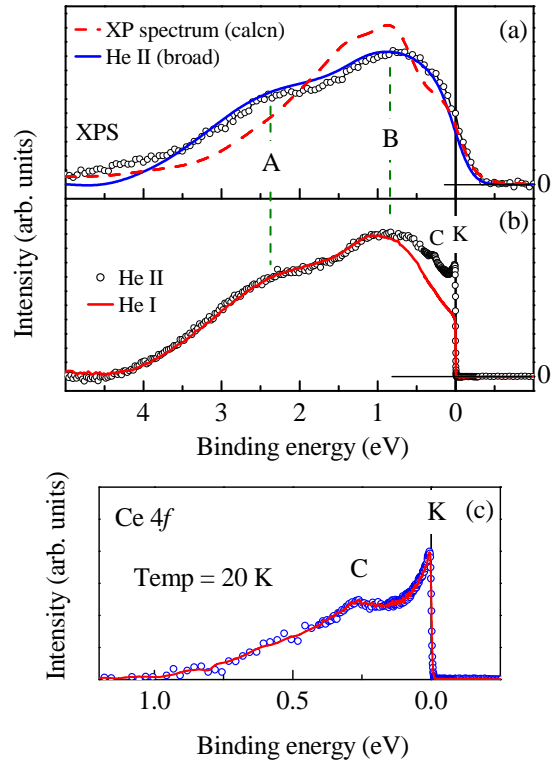}
\end{center}
\vspace{-8ex}
 \caption{Valence band spectra collected using (a) Al $K\alpha$
(XPS), and (b) He {\scriptsize II} and He {\scriptsize I} excitation
energies. The solid line in (a) represents the broadened He
{\scriptsize II} spectrum to take into account the energy resolution
corresponding to XP spectrum. Dashed line represent the calculated
XP spectrum. (c) The Ce 4$f$ spectral function obtained by
subtracting He {\scriptsize I} spectrum from the He {\scriptsize II}
spectrum.}
\end{figure}

The valence band in Ce$_2$CoSi$_3$ consists of Ce 5$d$, Ce 4$f$, Co
3$d$ and Si 3$p$ electronic states. Since the transition probability
of the photoelectrons in the photo-excitation process strongly
depends on the excitation energies, a comparison of the
photoemission spectra collected at different excitation energies
would help to identify experimentally the character of various
features constituting the valence band. In Fig. 1(a), we show the
valence band spectra collected at 20 K using Al K$\alpha$ photon
energy and the He {\scriptsize II}$\alpha$ and He {\scriptsize
I}$\alpha$ spectra are shown in Fig. 1(b). Each spectrum exhibits
two distinct features, A and B at about 1 eV and 2.3 eV binding
energies, respectively. Interestingly, the relative intensity of
these features does not change with such a large change in photon
energies. This is unusual as the photoemission cross sections
corresponding to Ce 5$d$, Ce 4$f$, Co 3$d$ and Si 3$p$ electronic
states have significantly different excitation energy
dependence.\cite{Yeh}

To verify the change in lineshape with better clarity, we have
broadened the He {\scriptsize II} spectrum by convoluting a Gaussian
of full width at half maximum (FWHM) = 0.4 eV to make the resolution
broadening comparable to that of the XP spectrum. The broadened
spectrum is shown by solid line superimposed over the XP spectrum in
Fig. 1(a). The lineshapes of both the spectra are almost identical.
In Fig. 1(b), the He {\scriptsize I} and He {\scriptsize II} spectra
are superimposed over each other to investigate the change in
lineshape when the energy resolution broadening is the same
($\sim$~5~meV). The spectra in the binding energy range beyond 1 eV
are found to be almost identical. All these observations suggest
that the binding energy range $>$ 1 eV of the valence band is
dominantly contributed by one kind of electronic states.

The intensity close to the Fermi level, $\epsilon_F$, exhibit
significantly different behavior as a function of excitation energy
(see Fig. 1(b)). He {\scriptsize II} spectrum exhibits large
intensity and two distinct features C and K near $\epsilon_F$, which
are not visible in the other spectra. The high energy resolution
employed in the He {\scriptsize I} and He {\scriptsize II}
measurements helped to resolve distinct signature of the features in
the vicinity of $\epsilon_F$. The photoemission cross section for Ce
4$f$ states at He {\scriptsize II} excitation energy is about 3
times larger than that at He {\scriptsize I} photon energy while it
is almost the same for Si 3$p$ states and double for Co 3$d$
states.\cite{Yeh} Thus, the features C and K can be attributed to
the photoemission signal primarily from the Ce 4$f$ states. We have
subtracted the He {\scriptsize I} spectrum from the He {\scriptsize
II} spectrum to delineate the Ce 4$f$ contributions. The subtracted
spectrum representing Ce 4$f$ band is shown in Fig. 1(c). The
distinct features, C and K corresponding to the spin orbit satellite
of the Abrikosov-Suhl resonance (ASR) and the main peak of ASR,
respectively, could clearly be identified.\cite{Ehm}

\begin{figure}
 \vspace{-2ex}
 \begin{center}
\includegraphics [scale=0.45]{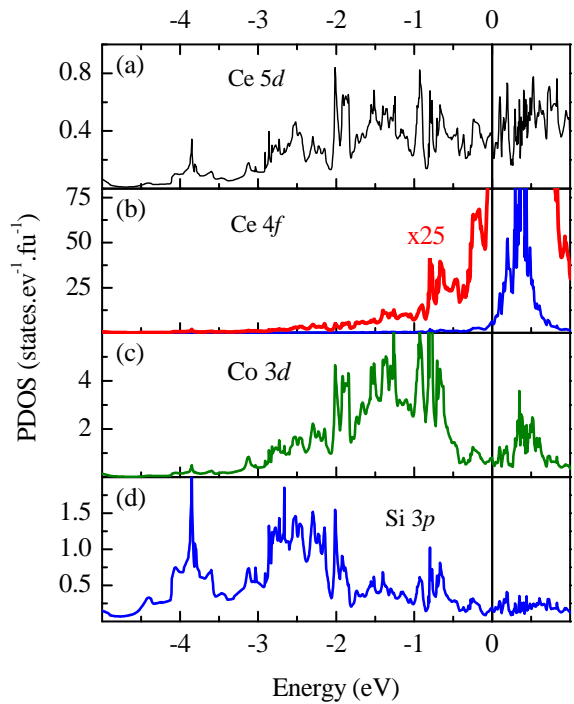}
\end{center}
 \vspace{-12ex}
 \caption{Calculated (a) Ce 5$d$  partial density of states (PDOS),
 (b) Ce 4$f$ PDOS, (c) Co 3$d$ PDOS and (d) Si 3$p$ PDOS. The thick solid
line in (b) represents the Ce 4$f$ PDOS rescaled by 25 times to show
the weak intensities at lower energies.}
\end{figure}

In order to verify the character of the features theoretically, we
have calculated the electronic band structure using FLAPW method.
The calculated partial density of states (PDOS) are shown in Fig. 2.
The dominant contribution in this energy range arises from the Ce
5$d$, Ce 4$f$, Co 3$d$ and Si 3$p$ PDOS as shown in Fig. 2(a), 2(b),
2(c) and 2(d) respectively. All the other contributions are
negligible in this energy range. Evidently, Ce 5$d$ contributions
are small and almost equally distributed over the whole energy range
shown. Ce 4$f$ band is intense and narrow as expected. In order to
provide clarity, we have rescaled the Ce 4$f$ partial density of
states (PDOS) by 25 times and shown by thick solid line in Fig.
2(b). Clearly, the 4$f$ PDOS contribute essentially in the vicinity
of the Fermi level ($<$ 1 eV binding energy). The intensity of Ce
4$f$ band is significantly weak at higher binding energies. This is
consistent with the observation in Fig. 1(b). Co 3$d$ states also
have finite contributions in this energy range due to the
hybridization between Co 3$d$ and Ce 4$f$ states.

Co 3$d$ and Si 3$p$ electronic states also exhibit significant
hybridization effect and appears dominantly in the binding energy
range larger than 0.5 eV. The bonding states contribute in the
energy range higher than 2 eV binding energy, where the Si 3$p$ PDOS
has large contributions. The antibonding features appear in the
energy range 0.5 to 2 eV, where Co 3$d$ contributions are dominant.
This suggests that the feature B in Fig. 1 has dominant Co 3$d$
character and the intensities corresponding to Si 3$p$ photoemission
contribute to feature A. The dominance of Co 3$d$ contributions in
this whole energy range shown presumably leads to unchanged spectral
lineshape with the change in photon energy as observed in Fig 1(a)
and 1(b).

In order to compare the experimental spectrum with the calculated
results, we have calculated the XP spectrum in the following way:
the Ce 5$d$, Ce 4$f$, Co 3$d$ and Si 3$p$ PDOS per formula unit were
multiplied by the corresponding photoemission cross sections at Al
$K\alpha$ energy. The sum of all these contributions was convoluted
by the Fermi distribution function at 20 K and broadened by the
Lorentzian function to account for the photo-hole lifetime
broadening. The resolution broadening is introduced via further
broadening of the spectrum by a gaussian function of FWHM = 0.4 eV.
The calculated spectrum is shown by dashed line in Fig. 1(a) after
normalizing by the total integrated area under the curve. The
intensities near $\epsilon_F$ in the experimental spectrum appears
to be captured reasonably well in the calculated spectra.

In the higher binding energy region; the intensity around 1 eV is
overestimated and that around 2.5 eV is underestimated. Since this
energy range contains dominant contribution from the Co 3$d$ states,
it naturally indicates that Co 3$d$ PDOS region is not well
described and correlations among Co 3$d$ electrons may be important
in determining the electronic structure in this energy range. In
order to verify this, we have calculated the electronic density of
states considering finite electron correlation, $U_{dd}$, among Co
3$d$ electrons. The spectral functions corresponding to different
$U_{dd}$ values are calculated from the LDA+$U$ results following
the procedure described above.

\begin{figure}
 \vspace{-2ex}
 \begin{center}
\includegraphics [scale=0.45]{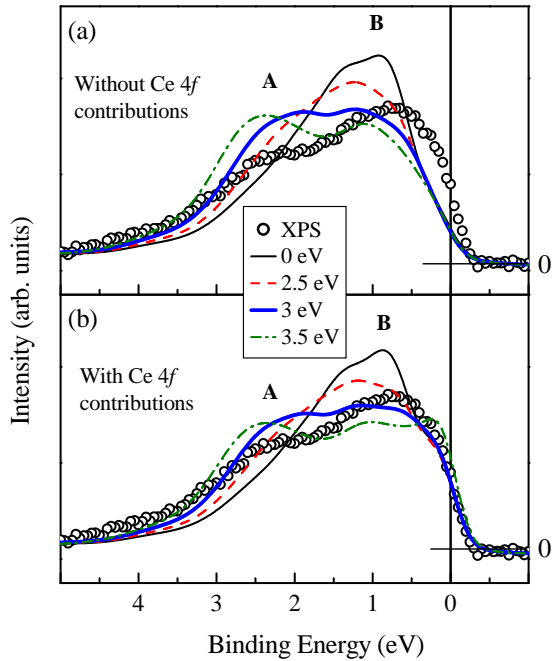}
\end{center}
 \vspace{-16ex}
 \caption{X-ray photoemission valence band spectrum (open circles)
is compared with the calculated spectral functions corresponding to
different $U_{dd}$-values. (a) The calculated spectral functions
without Ce 4$f$ contributions. (b) The calculated spectral functions
contains Ce 4$f$ contributions. Clearly, the results in (b) provide
better description than that in (a) revealing signature of Ce 4$f$
contributions in the vicinity of the Fermi level.}
\end{figure}

The lines in Fig. 3 represent the spectral functions for different
$U_{dd}$ values which are superimposed on the experimental XP
spectrum represented by open circles. In Fig. 3(a) we show the
calculated spectra without Ce 4$f$ contributions and the ones
including Ce 4$f$ contributions are shown in Fig. 3(b). It is
evident from the figure that the signature of the feature around 2.3
eV (feature B) becomes more and more prominent with the increase in
$U_{dd}$. Subsequently, the intensity of feature A reduces. Thus,
the feature B can be attributed to the photoemission signal from
electron correlation induced Co 3$d$ bands (lower Hubbard band) in
addition to the Si 3$p$ contributions. It is clear that the
calculated spectral functions corresponding to $U_{dd}$ $\sim$ 3 eV
is closest to the experimental spectrum compared to all other cases.
For higher values of $U_{dd}$, the correlation induced feature
becomes stronger and appears at higher binding energies.

\begin{figure}
 \vspace{-2ex}
 \begin{center}
\includegraphics [scale=0.45]{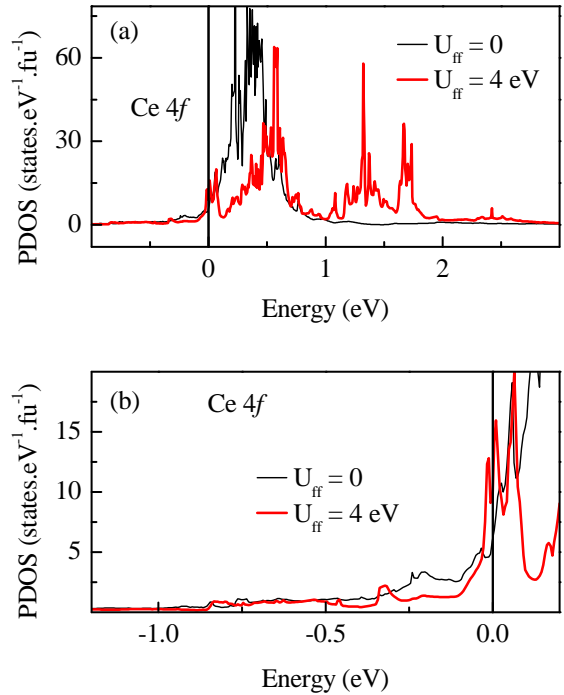}
 \end{center}
\vspace{-16ex}
 \caption{Calculated Ce 4$f$ partial density of states for different
electron correlation strength, $U_{ff}$ among Ce 4$f$ electrons. The
results corresponding to the whole energy range is shown in (a) and
that near Fermi level is shown in (b).}
\end{figure}

The comparison of Fig. 3(a) and 3(b) establishes that the
intensities near $\epsilon_F$ essentially arise due to the
photoemission from the occupied part of the Ce 4$f$ band. Co 3$d$
contributions are small in this energy range and the Co 3$d$
correlation induced effects has negligible influence on the spectral
intensity at $\epsilon_F$. Consideration of Ce 4$f$ bands in the
spectral function calculation leads to a better description of the
spectral intensities at the Fermi level. In order to investigate the
role of correlation among Ce 4$f$ electrons in the electronic
structure, we show the Ce 4$f$ PDOS calculated for $U_{ff}$ = 0 and
4 eV in Fig. 4. The unoccupied part of the spectral function
exhibits large redistribution. The occupied part exhibits small
change in lineshape as shown with better clarity in Fig. 4(b). The
peak at $\epsilon_F$ enhances and subsequently, the intensity
between 0 and -0.3 eV energies reduces. The rest of the occupied
part remains almost unchanged.

\begin{figure}
 \vspace{-2ex}
 \begin{center}
\includegraphics [scale=0.45]{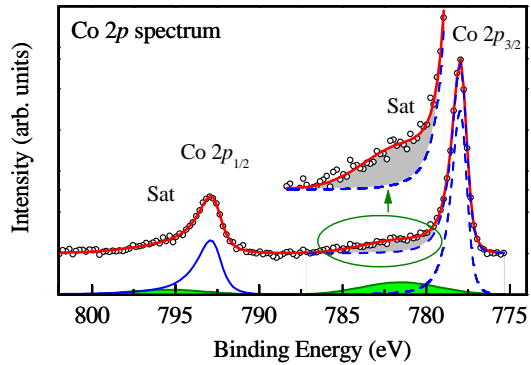}
 \end{center}
\vspace{-52ex}
 \caption{Co 2$p$ spectrum (open circles) exhibiting distinct signature
of main and satellite peaks. The satellite intensity is shown by
shaded region. Solid line passing through the experimental data
represents the fit comprising of a main peak (dashed line) and a
satellite (shaded peak).}
\end{figure}

The signature of electron correlations can also be observed in the
Co 2$p$ core level spectra.\cite{fujimoriRMP} In Fig. 5, we show the
Co 2$p$ spectrum collected at 20 K using Al $K\alpha$ radiation. The
spectrum consists of two spin orbit split features Co 2$p_{3/2}$ and
Co 2$p_{1/2}$ at 778 eV and 792.9 eV binding energies, respectively
(energy separation of about 14.9 eV). These binding energies are
identical to those found in elemental Co
metals.\cite{Nath,Schneider} This indicates that the valence state
of Co is very similar to the elemental Co metals. In addition, every
spin orbit split feature exhibits a weak but distinct shoulder at
higher binding energies. This has been shown more clearly by
rescaling and shifting this energy region in the same figure. The
energy separation between the main peak and the satellite is about 4
eV. Finite intensity of the satellite feature again establishes the
presence of electronic correlations among Co 3$d$
electrons.\cite{fujimoriRMP} Thus, our results establish that the
electron correlation strength among Co 3$d$ electrons are
significant and needs consideration to describe the electronic
structure of these systems.

\section{conclusions}

In summary, we have studied the electronic structure of
Ce$_2$CoSi$_3$ using high resolution photoemission spectroscopy and
{\it ab initio} band structure calculations. The experimental
results indicate the dominance of Co 3$d$ contributions in the
valence band. Si 3$p$ states appear at higher binding energies (2 -
3 eV). The Ce 4$f$ contributions appear essentially in the vicinity
of the Fermi level. High resolution employed in this study helped to
probe the Kondo resonance feature appearing at the Fermi level.
Although the contribution of the Co 3$d$ states at $\epsilon_F$ is
weak, Co 3$d$ states are found to be hybridized with the Ce 4$f$
states. The comparison of the experimental results with the
calculated ones indicate that Co 3$d$ electrons are strongly
correlated. These results thus, establish that in addition to the
electron correlations among Ce 4$f$ electrons, correlation among $d$
electrons needs to be considered to derive the electronic properties
of these systems.

\section{Acknowledgements}

One of the authors S.P., thanks the Council of Scientific and
Industrial Research, Government of India for financial support.

\end{document}